\documentclass{aa}  

\usepackage{graphicx}
\usepackage{txfonts}
\usepackage{xcolor}
\usepackage{orcidlink}
\usepackage{ulem}
\usepackage{hyperref} 
\usepackage{multirow}
\usepackage[switch]{lineno}

\def\apjl {ApJL}
\def\apjs {ApJS}

\def\aap {A\&A}

\def\R200 {R_{200}}

\definecolor{sele}{RGB}{255,105,180}  
\definecolor{facu}{RGB}{0,128,0}     
\definecolor{julian}{RGB}{0,0,255}  
\definecolor{vale}{rgb}{0.8,0,0}      
\definecolor{hernan}{RGB}{138,43,226}

\begin{document} 

\authorrunning{S. Levis et al.}

   \title{Cluster-green galaxy correlations: where do these galaxies live?}

  \author{Selene Levis\inst{1,2}\orcidlink{0000-0003-1887-776X},
        Facundo Rodriguez\inst{1,3}\orcidlink{0000-0002-2039-4372},
        Héctor J. Martínez\inst{1,3}\orcidlink{0000-0003-0477-5412},
        Valeria Coenda\inst{1,3}\orcidlink{0000-0001-5262-3822}
        \& Hernán Muriel\inst{1,3}\orcidlink{0000-0002-7305-9500}
   }

\institute{
        Instituto de Astronomía Teórica y Experimental, CONICET - UNC, Laprida 854, X5000BGR, Córdoba, Argentina
        \and
        Facultad de Matemática, Astronomía, Física y Computación, Universidad Nacional de Córdoba, Av. Medina Allende s/n, X5000HUA, Córdoba, Argentina
        \and
        Observatorio Astronómico, Universidad Nacional de Córdoba, Laprida 854, X5000BGR, Córdoba, Argentina
        }

   \date{Received XXXX; accepted XXXX}

   \abstract
   {Green valley (GV) galaxies are thought to represent a transitional population between star-forming and quiescent systems.
   However, their spatial distribution relative to galaxy systems, such as galaxy groups and clusters, remains unclear, particularly in relation to the large-scale environmental influence on galaxy quenching.} 
   {We aim to determine whether GV galaxies preferentially inhabit specific environments within galaxy systems, and to explore the physical nature of their location in the outskirts of massive haloes.} 
   {We analyse the spatial distribution of GV galaxies using the cluster–galaxy cross-correlation function (CCF), based on two datasets: the hydrodynamical simulation Illustris TNG300-1 and observational data from the Sloan Digital Sky Survey (SDSS DR18), cross-matched with the MPA–JHU DR7 catalogue. Galaxy systems with $\log(M_{200}/M_{\odot}) \geq 13.5$ are used as cluster centres, while galaxies classified as blue, green, or red, based on their location in the $(u-r)$ versus stellar mass diagram, serve as tracers for the correlation analysis.} 
   {In TNG, GV galaxies show an increasing relative fraction with cluster-centric distance, peaking in the outskirts, particularly for low-mass galaxies and haloes. In some cases, the GV fraction exceeds that of red galaxies. SDSS data reveal qualitatively similar, but weaker trends, with the GV fraction remaining below that of red galaxies at all scales. Most GV galaxies in the outskirts are satellites bound to the central FoF group, consistent with a backsplash or infall origin. Mock catalogues built from TNG and matched to SDSS selection functions reproduce the observational signal, indicating that projection effects drive the differences between datasets.}
   {GV galaxies preferentially reside in the outskirts of galaxy systems as satellites bound to the central halo, supporting a scenario in which they are transitioning objects influenced by environmental quenching.}

   \keywords{Galaxies: groups: general -- Galaxies: halos -- Galaxies: star formation --
   Galaxies: statistics -- large-scale structure of Universe
   }

\maketitle

\section{Introduction}\label{sec:intro}

Galaxies in the Universe are broadly found occupying two distinct sequences in the colour absolute magnitude (or colour vs. stellar mass diagram) (CMD): the blue cloud (BC), consisting of actively star-forming galaxies, and the red sequence (RS), composed of passive systems \citep{Strateva:2001}. This bimodality is well described by fitting the observed colour distribution at fixed absolute magnitude with the sum of two Gaussian functions, representing the blue and red populations, respectively \citep{Baldry:2004}. Blue galaxies typically exhibit late-type morphologies (spiral and irregular), whereas red galaxies are generally associated with early-type systems (elliptical and lenticular, \citealt{Driver:2006}). However, incorporating near-ultraviolet (NUV) data from the GALEX satellite \citet{Wyder:2007} revealed a more complex structure in the CMD. They identified an excess of galaxies in the region between the BC and the RS, defining what is known as the green valley (GV), a transitional zone occupied by galaxies with declining star formation. These galaxies exhibit lower specific star formation rates (sSFR) than BC galaxies of similar stellar mass, suggesting that they are in the process of quenching. Different criteria have been used to define the GV boundaries, including empirical lines in the colour mass diagram \citep{Schawinski:2014, Coenda:2018}, density-based colour ranges \citep{Bremer:2018}, the 4000 Å break \citep{Angthopo:2019}, and the shape of the sSFR distribution \citep{Estrada-Carpenter:2023}. As noted by \citet{Salim:2014}, GV galaxies are particularly valuable for studying quenching mechanisms, as their position in the CMD reflects an evolutionary path from the BC towards the RS.

Observational studies based on samples from the Sloan Digital Sky Survey (SDSS, \citealt{York:2000}), namely the Mapping Nearby Galaxies at APO project (MaNGA, \citealt{MANGA}), the SAMI Galaxy Survey \citep{Bryant:2015}, and the Galaxy And Mass Assembly survey (GAMA, \citealt{Baldry:2010}) show that GV galaxies exhibit physical properties that are intermediate between those of blue, star-forming galaxies and red, quiescent ones. Specifically, GV galaxies are predominantly early-type spirals and lenticulars \citep{Bait:2017, Mishra:2018}. Furthermore, the transition from the blue to the red sequence appears to correlate with a gradual weakening of the spiral arm structure \citep{Smith:2022}, suggesting a continuous morphological evolution. However, \citet{Schawinski:2014} showed that the GV does not represent a single evolutionary pathway, but rather comprises two distinct quenching modes: slow quenching in late-type disk galaxies, driven by gradual gas depletion over timescales of several Gyr, and rapid quenching in early-type systems associated with major mergers and morphological transformation. An analysis of radial profiles of the sSFR reveals that GV galaxies show suppressed star formation compared to star-forming galaxies of similar mass, with this suppression evident out to two effective radius \citep{Belfiore:2018}. Spatially resolved observations from SAMI reveal that galaxies in denser environments show a decline in sSFR from the outside-in, supporting an environmental cause for quenching such as ram pressure stripping or galaxy interactions \citep{Medling:2018}. The dispersion in sSFR along the star-forming sequence exhibits a characteristic U-shaped trend with stellar mass, attributed to stellar feedback at low masses and AGN feedback at high masses \citep{Davies:2019}. This implies that the quenching mechanism affects the galaxy as a whole, rather than being confined to an expanding central region. The relative abundance of GV galaxies suggests that typical crossing timescales in the CDM of 1–2 Gyrs, and this transition is independent of the environment \citep{Bremer:2018}. These findings support the notion that GV galaxies—mainly early-type disk systems—experience a sustained decline in gas supply, insufficient to maintain previous star formation levels, ultimately leading them to adopt passive colors. The cessation of star formation may be driven by a combination of internal processes and external factors. In this regard, \citet{Das:2021} report that the fraction of green galaxies remains consistently around 10–20\% regardless of environment, with approximately 10\% of these hosting an active galactic nucleus (AGN). The environmental effects appear to further influence the quenching process. \citet{Coenda:2018} find that quenching is more efficient in high-density environments, where both the quenching timescale and the star formation rates are lower. Studies of group environments show that the fraction of passive galaxies increases towards halo centres, while star-forming galaxies dominate the infall regions,  with a modest ($\sim$20\%) suppression of SFR already evident in galaxies within low-mass groups \citep{Barsanti:2018}. In cluster environments, galaxies with recently quenched star formation, identified through strong Balmer absorption, bare significantly more common than in the field and preferentially located in cluster outskirts, suggesting outside-in quenching on timescales of $\sim$1.5 Gyr \citep{Owers:2019}. Kinematic observations from SAMI show that gas-disturbed galaxies are most frequent in the infall regions of clusters, where ram pressure stripping during pericentric passage drives centrally concentrated star formation and eventual quenching \citep{Bagge:2025}. For lenticular galaxies, the dominant formation channel appears to be the fading of spirals in dense environments, though mergers contribute in lower-density settings \citep{Coccato:2022}. According to \citet{Sampaio:2024}, the environment significantly impacts galaxy morphology, particularly in low-mass systems. In summary, transition galaxies carry signatures of both secular evolution and environment-driven quenching.

Regarding the spatial distribution of galaxies, one of the most important findings from large-scale surveys is the strong dependence of galaxy clustering on colour. Early analyses using data from the SDSS revealed that red galaxies exhibit significantly higher clustering amplitudes and steeper correlation functions than blue galaxies \citealt{Zehavi:2002}). These differences were interpreted as a manifestation of the morphology–density relation \citep{Dressler:1980}, with red, early-type galaxies preferentially residing in denser environments and blue, and star-forming galaxies inhabiting lesser dense regions. Later work confirmed that this colour dependence holds across luminosity bins and is not reducible to luminosity-related effects alone (\citealt{Zehavi:2005}; \citealt{Zehavi:2011}).

Several studies have extended the clustering analyses to the GV. \citet{Coil:2008}, analysing galaxies at $z\sim1$, show that GV galaxies exhibit large-scale clustering comparable to that of red galaxies, yet their small-scale correlation slope and infall kinematics more closely resemble those of blue galaxies. This implies that a green galaxy is more likely to have a nearby red neighbor than a blue one. This dual behaviour suggests that green galaxies inhabit haloes similar in mass to those of red galaxies, but are more likely to reside in their outskirts rather than near the centres. The one-halo term observed for green galaxies compared to red ones supports this picture, indicating a lower radial concentration within haloes, possibly consistent with a scenario in which green galaxies migrate towards halo centres, becoming redder in the process.

Using SDSS and GALEX data, \citet{Heinis:2009} report that clustering amplitude increases monotonically with redder UV–optical colours, underscoring its sensitivity to recent star formation history. Further support for the transitional nature of GV galaxies comes from halo modeling: \citet{Krause:2013} find that green galaxies occupy haloes with intermediate masses and satellite fractions relative to blue and red populations. This intermediate status is also evident at higher redshifts. \citet{Lin:2019}, studying galaxies in a redshift range $0.5\lesssim z\lesssim2.5$, show a redshift-dependent trend: at $z>1.5$, green galaxies resemble blue galaxies more closely in their clustering, while at $z<1.5$ they are more similar to red galaxies. Furthermore, these authors report that GV galaxies exhibit a clustering amplitude comparable to that of AGN at similar redshifts, pointing to a potentially significant role of feedback processes in driving galaxy colour transformations.

The broader picture emerging from multiple studies suggests that blue galaxies are typically central objects in low-mass haloes, while red and green galaxies are more frequently satellites residing in virialised environments (e.g., \citealt{Yeong-Shang:2010}; \citealt{Krause:2013}). These results support a model in which the blue-to-red transition via the GV reflects not only internal processes—such as gas depletion and stellar ageing—but also environmental evolution, as galaxies migrate from filamentary structures into groups or clusters.

More recently, \citet{McNab:2021} investigate the abundance of transition galaxies in clusters, reporting that GV galaxies do not exhibit a significant excess in dense environments at high stellar masses, although a modest enhancement is observed at lower stellar masses.  Additionally, \citet{Joy:2025} employed marked correlation functions to assess the dependence of galaxy properties on the environment, finding that colour is the strongest tracer of local overdensity among the parameters they probed.

However, these studies have primarily rely on the two-point autocorrelation function (ACF) for colour-selected galaxy samples, or on cross-correlation function (CCF) between galaxy populations distinguished by colour. To date, only one recent study explore the dependence of galaxy–cluster clustering on galaxy properties via cluster–galaxy cross-correlations. At low redshift, \citet{Comparat:2025} analyse correlations between X-ray selected clusters and colour-defined galaxy samples, finding that red galaxies dominate the cluster–galaxy cross-correlation signal, while blue galaxies show a clear suppression.

Together, these results trace a coherent picture: galaxy colour encodes information about environment, and the transition from blue to red is accompanied by systematic changes in spatial distribution. GV galaxies, in particular, exhibit clustering signals that reflect both their intermediate star-forming status and their emerging association with denser structures.

\begin{figure}[!h]
\centering
{\includegraphics[width=0.45\textwidth]{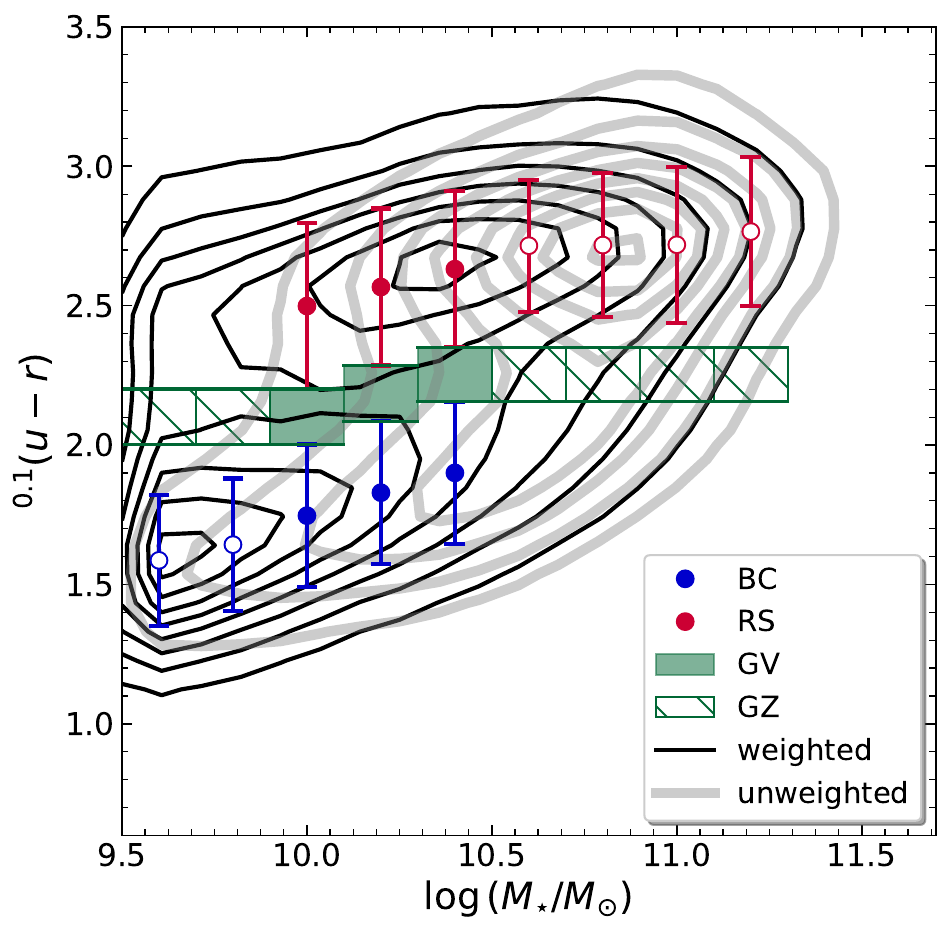}
\caption{\label{fig:cdm} $^{0.1}(u-r) $ colour-stellar mass diagram (CMD) for MGS DR18 galaxies restricted to $0.025\leq z \leq 0.15$ and to $\log(M_{\star}/M_{\odot}) \geq 9.5$. Solid blue and red circles indicate the centres of the Gaussian components that best fit the blue and red populations, respectively. Open symbols of the same colours mark cases where only the blue cloud or red sequence is present. The shaded green region indicates the green valley (GV), while the hatched area marks the green zone (GZ). Isocontours of galaxy number density are shown as black lines for the weighted distribution and grey lines for the unweighted case.}}
\end{figure}

In a recent study, \citet{Levis:2025} examine the evolution of various properties of galaxies, in and around groups, using the IllustrisTNG simulations \citep{Nelson:2019}, classifying them according to their trajectories. They find that backsplash galaxies exhibit the highest fraction of green galaxies.

The two-point correlation function is a powerful statistical tool for quantifying the spatial distribution of galaxies. In this work, we investigate the colour dependence of galaxy clustering by analysing the cross-correlation between galaxy systems (from groups to clusters) and galaxies, with a particular focus on the GV population, in order to understand the spatial locations these galaxies occupy in the vicinity of systems.

The article is structured as follows. Section \ref{sec:sample} introduces the clusters and galaxies samples used in this study, drawn from the IllustrisTNG simulations and SDSS observations. Section \ref{sec:corrfunc} outlines the methodology, focusing on the use of the cross-correlation function. In Sect. \ref{sec:results}, we present the results from the simulations and compare them with the observational data. Furthermore, it provides a comparison with the mock catalog to investigate discrepancies between the simulations and observations. Finally, Sect. \ref{sec:conclusions} summarises the conclusions of the work. 

\section{The samples}\label{sec:sample}

\subsection{Simulations IllustrisTNG} \label{sec:sample_simus}

From a theoretical perspective, we perform our analysis using  the TNG300-1 run of the IllustrisTNG project \citep{Nelson:2019}, a cosmological gravo-magnetohydrodynamical simulation that evolves the formation of structure from $z = 127$ to the present day under a $\Lambda$CDM cosmology consistent with \citet{PlanckColab:2016}. The adopted cosmological parameters are $\Omega_{\Lambda,0}=0.6911$, $\Omega_{m,0}=0.3089$, $\Omega_{b,0}=0.0486$, $\sigma_{8}=0.8159$, $n_{s}=0.9667$, and $h=0.6774$. This simulation models a $300\,\mathrm{Mpc}^3$ comoving volume and incorporates a comprehensive treatment of galaxy formation processes, including gas dynamics, star formation, feedback, and chemical enrichment.

To trace the galaxy population, we select 175056 subhaloes from the Subfind catalogue \citep{Springel:2001} with stellar masses $\log(M_{\star}/M_{\odot}) \geq 9.5$. We assign each galaxy rest-frame absolute magnitudes in the SDSS $u-$ and $r-$bands using synthetic photometry from the IllustrisTNG Supplementary Data Catalogs \citep{Nelson:2018}, which include dust attenuation effects. This galaxy sample serves as the tracer population in our cluster–galaxy cross-correlation analysis.

Many authors have used the colour–stellar mass plane (e.g. \citealt{Wyder:2007, Brammer:2009, Walker:2013, Lee:2015, Trayford:2015, Coenda:2018, Eales:2018, Phillipps:2019, Parente:2025}) to define the GV population, adopting different criteria to delineate the boundaries between the blue cloud and the red sequence.
In this work, we adopt the definition proposed by \citet{Levis:2025}, who identify red, blue, and green galaxy populations in the $(u - r)$–stellar mass diagram (see their Fig. 1 for further details). Their classification is based on modelling the colour distribution within stellar mass bins using single or double Gaussian components, selected according to the Bayesian Information Criterion (BIC). From these fits, they define the GV as the region where both red and blue populations coexist, and extend this definition towards higher stellar masses where either the red sequence or the blue cloud remains but exhibits a broadened colour distribution. Following their scheme, we classify all galaxies within both the GV and the extended green zone (GZ) as green galaxies.
Applying this criterion to our sample yields $92798$ blue galaxies, $29477$ green galaxies, and $52781$ red galaxies.

Furthermore, we extract $1146$ galaxy systems from the friends-of-friends (FoF) group catalogue at 
$z=0$. The TNG FoF catalogues are generated from the
simulations using the standard FoF algorithm \citep{Springel:2001}. This method links together particles separated by less than a fixed fraction of the mean interparticle spacing (typically $0.2$), thereby 
identifying groups that correspond approximately to virialized dark matter haloes. Within each FoF halo, 
the most bound particle is taken as a reference centre.
We select those with masses $\log(M_{200}/M_{\odot}) \geq 13.5$. Here, $M_{200}$ refers to the mass enclosed within a radius where the average density is 200 times the critical density of the Universe at the same redshift. This mass threshold ensures that each system hosts at least five satellite galaxies, providing reliable statistics for clustering measurements.

\subsection{Observational Data} \label{sec:sample_obs}

\subsubsection{The galaxy sample}

We draw our observational galaxy sample from the Main Galaxy Sample (MGS; \citealt{Strauss:2002}) in SDSS Data Release 18 (DR18; \citealt{Almeida:2023}). We restrict our selection to the contiguous Northern Galactic Cap Legacy footprint $(\sim7500\,\mathrm{deg}^2)$ within $0.025 \leq z \leq 0.15$. To minimise incompleteness at the bright end, we exclude all galaxies with $r \leq 14.5$, resulting in a final apparent‑magnitude interval $14.5 \leq r \leq 17.77$. 

We correct all apparent magnitudes for Galactic extinction using the maps of \citet{sch98}, then compute absolute magnitudes by applying K‑corrections and shifting to $z=0.1$ with {\sc Kcorrect} v4.1 \citep{Blanton:2003}. All magnitudes throughout are quoted in the AB system.  We further require $\log(M_{\star}/M_{\odot}) \geq 9.5$, drawing stellar masses from the MPA–JHU value‑added catalogue based on SDSS‑DR7 \citep{Abazajian:2009} through a direct cross‑match with our parent sample. We adopt the same cosmological parameters for the observational data as those used in the TNG simulations.

Figure \ref{fig:cdm} shows the colour–magnitude diagram (CMD) for our sample of $409348$ galaxies. Due to the flux limit of the parent catalogue, each object is assigned a $1/V_{\max}$ weight  \citep{Schmidt:1968} in our CDM analysis. Galaxies are then split into blue, green and red populations, adopting exactly the same classification criteria as those applied to the TNG sample. Fig. \ref{fig:cdm} also shows the difference between the weighted and unweighted CMD. From this classification, we obtain $141050$ blue galaxies, $43046$ green galaxies, and $225252$ red galaxies.

We assessed the impact of stellar mass and photometric uncertainties on the classification of galaxies into blue, green, and red. Regarding the mass uncertainties, according to \citet{Kauff:2003}, the typical 95\% confidence interval in stellar mass corresponds to $\sim0.3$ dex in $\log(M_{\star}/M_{\odot})$, i.e. $\sigma \sim 0.15$ dex for Gaussian errors. We simulated this by adding a random Gaussian perturbation $\varepsilon$ to each stellar mass, obtaining $\log(M_{\star})_\mathrm{pert}=\log(M{\star}) + \varepsilon$. After reclassifying galaxies in the CMD, 99\% of blue, 95\% of green, and 100\% of red galaxies retained their original classification. Thus, typical stellar mass uncertainties do not significantly affect our colour-based analysis. We evaluated the impact of photometric errors and $k$-corrections on our classification through a Monte Carlo test. Since kcorrect does not provide individual $k$-correction errors, we estimated them using kcorrect as follows. Input magnitudes were perturbed with Gaussian noise according to their photometric errors, and $k$-corrections were recomputed for 30 realizations per galaxy. From the resulting $(u_\mathrm{pert}-r_\mathrm{pert}) + (k_\mathrm{pert}^u - k_\mathrm{pert}^r)$ distributions, we derived standard deviations adopted as individual $\sigma(u-r)$ values. We then generated perturbed colours $(u-r)_\mathrm{pert}$ in an analogous way as we did with stellar mass uncertainties and reclassified galaxies in the CMD. The comparison between original and perturbed samples shows that the fractions of blue, green, and red galaxies remain essentially unchanged. After reclassifying galaxies in the CMD, 94\% of blue, 82\% of green, and 100\% of red galaxies retained their original classification. Thus, uncertainties in the photometry should not significantly affect our results.

\subsubsection{The sample of galaxy systems}

To identify galaxy groups within the SDSS DR18, we employed a methodology outlined by \citet{Rodriguez:2020}. This approach combines the FoF method as in \citet{Merchan:2005} with halo-based techniques \citep{Yang:2005}. The process begins by identifying gravitationally bound systems using a percolation method, an adaptation of the technique described by \citet{Huchra:1982}. Each identified group is assumed to contain at least one luminous central galaxy, and its initial properties are estimated based on the collective luminosity of its constituent members. Subsequently, an iterative, halo-based refinement step is applied to enhance the reliability of group membership and recalculate halo properties. This iterative procedure, which continues until convergence is achieved, ensures accurate system identification across a wide spectrum of group sizes, from small associations to extensive clusters. The resulting galaxy group catalogue provides essential information, including galaxy membership, spatial coordinates, and estimations of halo mass and radius. The halo mass estimation, $M_{200}$, is derived using an abundance matching technique \citep{Vale:2004, Kravtsov:2004, Conroy:2006, Behroozi:2010}, which posits a direct correlation between a group's characteristic luminosity and its dark matter halo mass. We compute the radius, $R_{200}$, following equation 7 of \citet{Rodriguez:2020}. Furthermore, galaxies within each group are categorized as either central or satellite, with the brightest galaxy consistently designated as the central galaxy \citep{Rodriguez:2021,Rodriguez:2022}.

From the identified groups, we select only those with halo masses $\log(M_{200}/M_{\odot})\geq 13.5$ to focus on galaxy clusters. We identify the brightest galaxy in each group as the central and adopt its position as the group centre. This choice aligns with the established convention for central/satellite classification, where the most luminous member traces the gravitational potential of the host halo \citep{Rodriguez:2021, Rodriguez:2022}. Our final sample comprises $5612$ galaxy systems.

The group catalogue used in our analysis has been validated against weak-lensing masses \citep{Gonzalez:2021} and has been successfully used in a variety of studies of galaxy environments (e.g. \citealt{Alfaro:2022, Rodríguez-Medrano:2023, Parente:2025}). Furthermore, a recent comparative analysis by \citet{Kakos:2024} has shown that key conclusions regarding galaxy clustering and the galaxy-halo connection remain robust when using either the \citet{Yang:2012}, \citet{Tempel:2017}, or \citet{Rodriguez:2020} catalogues, giving us high confidence in the robustness of our results.

\subsection{Mock catalogue} \label{sec:sample_mock}

To bridge the gap between theoretical models and observations, we construct a mock catalogue as an intermediate stage 
by placing the observer at the origin of the TNG300 simulation box. We replicate the TNG300 volume as needed to match the spatial coverage of the SDSS DR18 spectroscopic survey, leveraging the periodic boundary conditions of the simulation. We calculate the redshift of each galaxy by combining its cosmological distance with the shift caused by peculiar velocities. By combining these redshifts with the intrinsic absolute magnitudes from TNG300, we derive the apparent magnitudes, allowing us to replicate the flux-limited selection of SDSS DR18 through the magnitude cutoff. Also, we construct an angular selection mask to mimic the DR18 footprint, following methodologies similar to those used by \citet{Rodriguez:2015}, but adjusting for the updated geometry and completeness of DR18. A crucial consideration is the classification of galaxies as centrals or satellites, which, in observational surveys like SDSS, depends on group-finding algorithms and introduces inherent uncertainties. To ensure consistency, we apply the group-finder algorithm from \citet{Rodriguez:2020} to our mock catalogue, mirroring the approach taken in observational data.

After applying the redshift range ($0.025 \leq z \leq 0.15$) and the SDSS flux limit ($14.5 \leq r < 17.77$), the final sample contains $128393$ galaxies with stellar masses $\log(M_{\star}/M_{\odot}) \geq 9.5$, and $9745$ galaxy systems with $\log(M_{200}/M_{\odot}) \geq 13.5$, used as tracers and cluster centres, respectively, in the projected CCF analysis. Among these galaxies, $26865$ are classified as blue, $19574$ as green, and $81954$ as red.

\section{Cluster-galaxy cross-correlation functions}\label{sec:corrfunc}

To investigate the spatial distribution of green galaxies, we utilised the cross-correlation function, which enables us to quantify differences in the clustering of distinct galaxy samples. For the IllustrisTNG data, we computed the 3D CCF between galaxy system centres (C) and galaxy tracers (G), following the classical estimator of \citet{Davis:1983}:

\begin{equation}
\xi(r) = \frac{N_R}{N_G}\frac{CG(r)}{CR(r)} - 1
\end{equation}
here $CG(r)$ and $CR(r)$ are the counts of system–galaxy and system–random pairs separated by a distance $r$, and $N_{G}$, $N_{R}$ are the numbers of galaxies and random points in the sample, respectively. The random catalogue (R) was constructed by uniformly distributing points within a box of the same size as the IllustrisTNG300 simulation volume.

To conduct an analysis of observations comparable to that of simulations and to avoid redshift-space distortion effects, we determine the two-coordinate cross-correlation function. Here, $r_l$ represents the dimension parallel to the line of sight, and $r_p$ represents the dimension perpendicular to it. Subsequently, we integrate over the $r_l$ direction to calculate the projected cross-correlation function,
$\xi (r_p)$: 

\begin{equation}\label{eq:wp}
\xi(r_p) = 2 \int_{r_{\min}}^{r_{\max}} \Xi(r_p, r_l)\,\mathrm{d}r_l
\end{equation}
where $r_{\min} = 0.02\,h^{-1}\mathrm{Mpc}$ and $r_{\max} = 30\,h^{-1}\mathrm{Mpc}$ 
and the estimator for $\Xi(r_p,r_l)$ is given by:

\begin{equation}
\Xi(r_p, r_l) = \frac{CG(r_p, r_l)}{CR(r_p, r_l)} \cdot \frac{N_{R}}{N_{G}} - 1
\end{equation}
where $CG(r_p, r_l)$ and $CR(r_p, r_l)$ represent the counts of cluster–galaxy and cluster–random pairs separated by $r_p$ and $r_l$, and $N_{G}$, $N_{R}$ are the numbers of galaxies and random points in the sample. The random catalogue was generated by uniformly distributing points across the SDSS footprint, reproducing the angular distribution, as well as the redshift distribution of the SDSS galaxy sample. For these computations we use the same correlation code used in \citet{Rodriguez:2022}.

\section{Results} \label{sec:results}

\begin{figure*}[!h]
\centering
{\includegraphics[width=0.9\textwidth]{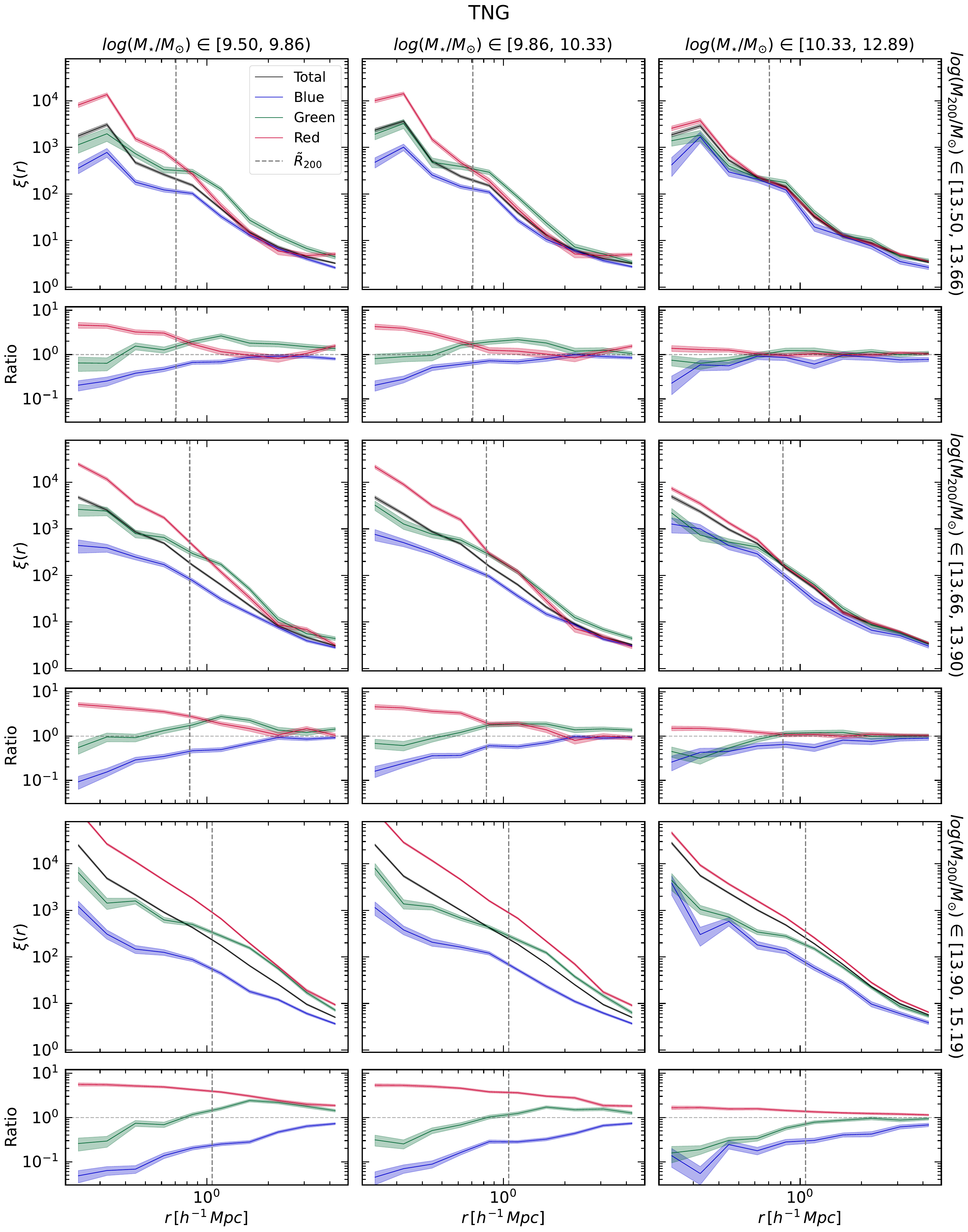}
\caption{\label{fig:ccf_tng} Upper panels: Cluster–galaxy cross-correlation functions (CCFs) derived from the IllustrisTNG300-1 simulation. The black curve shows the full galaxy sample; blue, green, and red curves correspond to the respective colour subsamples. Rows correspond to terciles of galaxy system $M_{200}$, and columns to terciles of galaxy stellar mass. Shaded regions show uncertainties estimated via jackknife resampling. Lower panels: Ratios between the CCFs of the red, green, and blue subsamples and that of the total sample. Grey dashed lines denote the median $R_{200}$ corresponding to each tercile of $M_{200}$.}} 
\end{figure*}

Figure \ref{fig:ccf_tng} shows the cluster–galaxy CCFs computed from the IllustrisTNG 300-1 simulation. The upper panels display the CCFs for the full galaxy sample (black curve) and for the colour-selected subsamples: blue, green, and red galaxies (blue, green, and red curves, respectively). The lower panels show the ratio of each subsample's CCF to that of the full sample, highlighting each subsample's unique cross-correlation characteristics. Rows correspond to increasing $M_{200}$ terciles (from top to bottom), and columns to increasing stellar mass terciles (from left to right). For each $M_{200}$ tercile, the median value of $r_{200}$ is indicated by a grey dashed line. Shaded regions represent jackknife-estimated uncertainties.

We observe that within the clusters, the CCF of green galaxies exhibits an intermediate behaviour between that of blue and red galaxies. In particular, for galaxies in the highest stellar mass range, the green and blue CCF become more similar, especially as the stellar mass decreases. When analysing the ratio between the CCF of the colour-defined subsamples and that of the complete sample, we find that the likeness of finding green galaxies increases from the central to the outer regions of galaxy systems and tends to be higher in the outskirts, particularly for galaxies with lower stellar mass. Specifically, in the lowest and the second $M_{200}$ tercile (for galaxies with low and intermediate stellar masses), green galaxies dominate in the outskirts, even surpassing red galaxies. In the remaining combinations, although the relative fraction of green galaxies also increases with distance from the centre, it does not exceed that of red galaxies.

\begin{figure*}[!h]
\centering
{\includegraphics[width=0.9\textwidth]{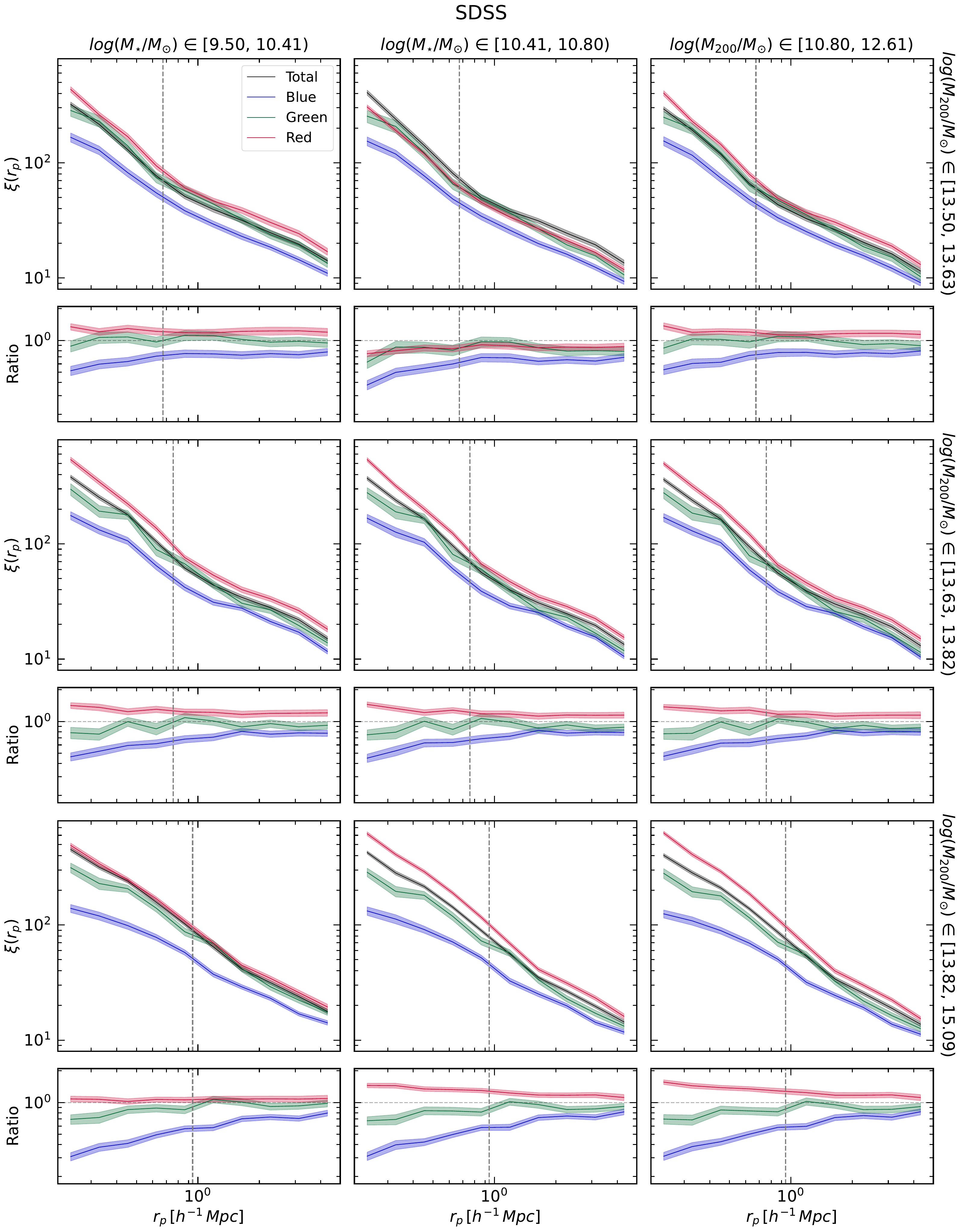}
\caption{\label{fig:ccf_sdss} CCFs and their ratios for the SDSS DR18 sample, analogous to Fig. \ref{fig:ccf_tng}. Rows correspond to terciles of galaxy system $M_{200}$, and columns to terciles of galaxy stellar mass. Shaded regions indicate uncertainties estimated via jackknife resampling. The grey dashed lines indicate the median $R_{200}$ for each of the respective tercile of $M_{200}$.}}
\end{figure*}

We perform a cross-correlation analysis using the SDSS sample to examine whether similar trends are observed in real data. 
Figure \ref{fig:ccf_sdss} presents the observational results in plots 
analogous to those in Fig.~\ref{fig:ccf_tng}. In general, the observational 
data show trends consistent with the simulation results. The relative 
fraction of red galaxies increases towards the centres of the systems, 
whereas the fraction of blue galaxies rises towards the outer regions but 
never surpasses that of red galaxies within the observed distance range. 
Green galaxies display an intermediate behaviour: within $R_{200}$ their 
distribution resembles that of blue galaxies, while in the outer regions it 
becomes more similar to that of red galaxies. Similarly as 
Fig.~\ref{fig:ccf_tng}, we note that for green galaxies the relative fraction 
increases towards the outskirts and the infall region, particularly in low 
halo-mass systems, where green galaxies tend to match (or even surpass, in 
the case of the simulations) the relative fraction of red galaxies.

\begin{figure*}[!h]
\centering
{\includegraphics[width=0.9\textwidth]{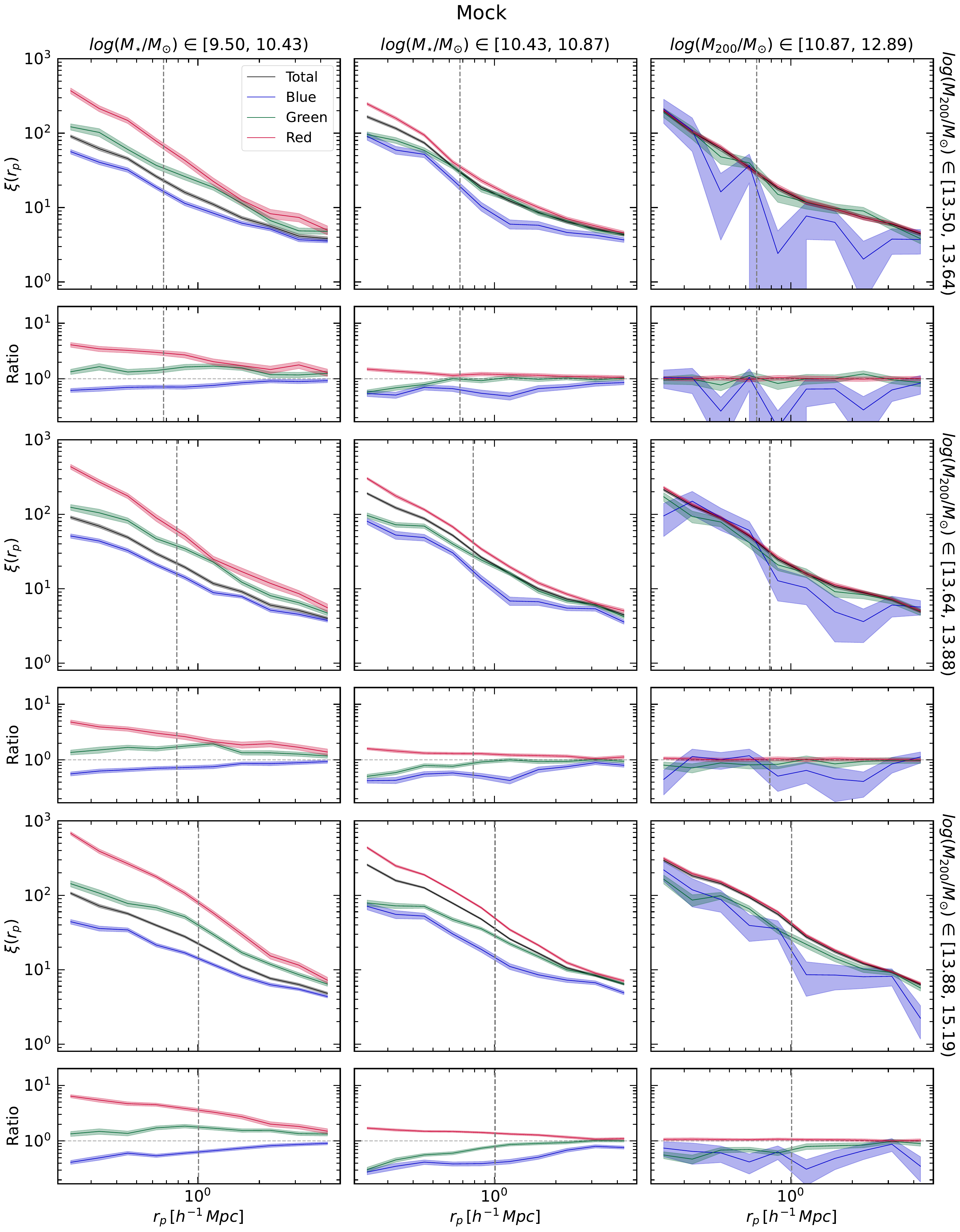}
\caption{\label{fig:ccf_mock} CCFs and their ratios for the mock sample, analogous to Fig. \ref{fig:ccf_tng}. Rows correspond to terciles of galaxy system $M_{200}$, and columns to terciles of galaxy stellar mass. Shaded regions indicate uncertainties estimated via jackknife resampling. The grey dashed lines indicate the median $R_{200}$ for each of the respective tercile of $M_{200}$.}}
\end{figure*}

In Sect. \ref{sec:sample_obs} we explored how colour uncertainties could affect our classification of green galaxies in the SDSS sample when colour are perturbed accordingly with expected errors in photometry and $k-$corrections. We computed the global CCF between the sample of colour-perturbed green galaxies and the groups in the SDSS. This is shown in Fig. \ref{fig:ccf-sdss-color-pert}, where it can be seen that, within errors, the resulting signal is consistent with that of the original sample. We therefore do not expect colour uncertainties to be an issue in the CCF of green galaxies and systems.

We explored whether the observed CCF of green galaxies is unique to them and not what we would expect from mixing up together blue and red galaxies. We selected a sample that includes all galaxies 0.1 dex above the GV/GZ and 0.1 dex below it, hereafter the (B+R) sample, and repeated the global CCF analysis. These are in Fig. \ref{fig:ccf-tng-ap} for TNG galaxies, and in Fig. \ref{fig:ccf-sdss-ap} for SDSS galaxies. It can be seen that in both cases, the CCF of the (B+R) sample presents a clearly distinct behaviour compared to green galaxies.

Previous studies suggest that most blue galaxies are typically central and isolated, whereas the fraction of satellites is higher among red and green galaxies, and are more commonly found in virialised structures (e.g., \citealt{Heinis:2009, Yeong-Shang:2010}). We compute the central and satellite fractions of green galaxies located in the outskirts of galaxy clusters, specifically between one and two $R_{200}$ from the group centre, for systems with $\log(M_{200}/M_{\odot}) \geq 13.5$ as well as for the three $M_{200}$ terciles, in both the TNG simulation and SDSS observations, as summarised in Table~\ref{tab:cen_sat}. The table also reports the fraction of those satellites that belong to the central FoF group. 

\begin{table}[ht]
\centering
\caption{
Satellite fractions of GV galaxies in cluster outskirts 
($1 \leq R/R_{200} \leq 2$) for TNG and SDSS.}
\renewcommand{\arraystretch}{1.3}
\begin{tabular}{l c c c c}
\hline
Sample & TNG & SDSS & TNG & SDSS \\
       & \multicolumn{2}{c}{[\%]} & \multicolumn{2}{c}{[\%] of central FoF} \\
\hline
   whole sample               & 88.87 & 78.66 & 99.17 & 89.38 \\
1\textsuperscript{st} tercile & 92.73 & 86.33 & 100.0 & 95.83 \\
2\textsuperscript{nd} tercile & 90.85 & 78.14 & 99.66 & 91.61 \\
3\textsuperscript{rd} tercile & 87.00 & 76.54 & 98.72 & 86.25 \\
\hline
\end{tabular}
\label{tab:cen_sat}
\end{table}

For the full sample, the fraction of centrals and satellites exhibits a qualitatively similar trend in both the simulation and observations: GV galaxies on the outskirts are predominantly satellites. In the SDSS data, only $\sim21\%$ are classified as centrals, while in the TNG simulation this fraction drops to $\sim11\%$. To determine whether these satellites are part of the central halo or belong to infalling structures, we analyse their group membership. In SDSS, approximately $89\%$ of the satellites in this radial range are gravitationally bound to the central FoF group whose outskirts they inhabit; in TNG, this fraction increases to $99\%$.

Additionally, we find that the proportion of GV galaxies in the outskirts that are satellites is higher in lower-mass clusters, in both the simulations and the observations. The fraction of these satellites that are gravitationally bound to the central FoF group also increases with decreasing cluster mass. In all cases, the corresponding fractions are slightly higher in the simulations than in the observations. The predominance of GV galaxies as satellites bound to the central FoF is consistent with the results of \citet{Levis:2025}, who, using a different methodology and a sample of group galaxies from the TNG simulations, report that most GV galaxies are backsplash galaxies, i.e. galaxies that have experienced a single pericentric passage and are therefore likely already bound to the system and undergoing quenching. An alternative scenario is that these galaxies are on their first infall into the central group.

To understand the origin of the discrepancies between the results from simulations and those derived from observational data, we perform an analogous analysis over the mock catalogue described in Sect. \ref{sec:sample_mock} which is shown in Fig. \ref{fig:ccf_mock}. Overall, the trends for green galaxies in the mock are more similar to SDSS. The relative rise of the correlation signal of green galaxies in the outskirts of systems seen in TNG vanishes in the mock. This behaviour indicates that the better part of the simulation–observation tension seen when comparing TNG and SDSS is naturally explained by projection effects, and the associated mixing of galaxies at different distances along the line of sight.

However, even when mock results are more similar to SDSS, the differences between green galaxies and the others are stronger in the mock. These differences between the mock and the observations may be due to a number of factors that prevents a one-to-one comparison. For instance the modelling of stellar populations in the simulation does not fully reproduce the observations (e.g. \citealt{Nelson:2019}). On the other hand, the biases introduced by the SDSS photometry measurement procedure are not fully accounted for in the simulation modelling. Furthermore, in the observations, the best way to determine the green valley in a colour-mass diagram, requires the use of colours not available in the TNG data.

\section{Conclusions}\label{sec:conclusions}

In this work, we investigate the spatial distribution of green valley (GV) galaxies in relation to galaxy systems using the cluster-galaxy cross-correlation function (CCF). Our main objective is to determine whether GV galaxies preferentially inhabit specific environments.

We employ two complementary datasets: the magneto-hydrodynamical simulation IllustrisTNG 300-1, and observational data from the Sloan Digital Sky Survey (SDSS-DR18), cross-matched with the DR7 MPA–JHU catalogue. In both samples, we use galaxies with $\log(M_{\star}/M_{\odot}) \geq 9.5$ at $z=0$ as tracers, and galaxy systems with $\log(M_{200}/M_{\odot}) \geq 13.5$ at $z=0$ as centres for the CCF. Galaxies were classified by colour (blue, green, red) based on their location in the $(u-r)$ colour–stellar mass diagram. We apply the appropriate correlation function for each dataset —3D for TNG and projected for SDSS— to perform the analysis. We recall that this comparison is qualitative in nature, aimed at identifying general trends rather than establishing a one-to-one correspondence between the simulations and observations.

The results from the IllustrisTNG simulation show a clear trend: the relative fraction of GV galaxies tends to increase towards the outer regions of galaxy systems and reaches a maximum in their outskirts, particularly for galaxies of lower stellar mass and in less massive systems. In these cases, the GV fraction in the outskirts even surpasses that of red galaxies. For more massive systems and more massive galaxies, although the effect is still present, it is less pronounced. The cross-correlation analysis based on SDSS data reveals qualitatively similar trends to those observed in the simulations, although with a weaker signal. Notably, in the SDSS data, the relative fraction of GV galaxies never exceeds that of red galaxies at any spatial scale. We examine the fraction of central and satellite GV galaxies in the outskirts of galaxy systems and find that the majority are satellites already gravitationally bound to the central system, residing in the peripheral regions around $R_{200}$. These results support the scenario in which GV galaxies are predominantly associated with backsplash or infalling populations likely undergoing environment-driven quenching. This interpretation is consistent with the findings of \citet{Levis:2025}, who, using an independent methodology, report that most backsplash galaxies are located in the GV.

We explore the causes of the differences in the signal between simulations and observations. We constructed a mock catalogue from the IllustrisTNG simulation data, applying the same selection functions and observational constraints as those used in the SDSS. The mock results closely reproduce the observational findings, suggesting that the discrepancy is not due to intrinsic differences between simulated and real galaxy populations, but rather to projection effects inherent to redshift-space observations. However, it is worth noting that cluster properties in the SDSS and mock samples are derived through the abundance matching method, and thus, they should be considered as proxies rather than direct measurements. The fact that the results from the mock sample reproduce the trends seen in real data supports two main conclusions: first, that the signal detected is genuine and robust; and second, that the identification of systems and the method for assigning cluster properties are not introducing significant biases relative to the fully three-dimensional information available in the simulation, beyond the expected limitations from projection and sample selection.

Taken together, these findings show the important role of environment in quenching star formation in transition galaxies. GV galaxies preferentially reside in the outskirts of galaxy systems and are predominantly satellite members of groups and clusters, highlighting that environment-driven processes are key to their evolution. Specifically, our results point out to:

\begin{enumerate}
    \item GV galaxies are found mostly in the outskirts of galaxy systems (in the regions around the virial radius, $R_{200}$), particularly for lower-mass galaxies and less massive haloes. This suggests that they may predominantly correspond to galaxies on their first infall into the system or to backsplash galaxies. This is in agreement with the results of \citet{Owers:2019}, who observationally find that recently quenched cluster galaxies are concentrated around $\sim0.5$–$1,R_{200}$, consistent with ram-pressure stripping quenching star formation on short ($\sim$1 Gyr) timescales.
    \item Most GV galaxies in these outskirts are satellites (either infalling newcomers or backsplash objects) already bound to the central system. This implies that many GV galaxies are in the process of quenching due to environmental effects during or shortly after their infall. Our results therefore support a scenario in which transitional GV galaxies are often on their first infall into a cluster or are recent departures that have passed through pericentre, linking their colour transformation to cluster-related processes. Consistently, cosmological simulations show that backsplash galaxies exhibit the highest fraction of green galaxies \citep{Levis:2025}.
    \item The excess of GV galaxies around clusters is a genuine environmental signal rather than an artifact of observation or simulation. The qualitative agreement between the IllustrisTNG predictions and SDSS measurements, together with the mock catalogue test, demonstrates that the spatial trend we detect is robust.
\end{enumerate}

In summary, the detection of a relative excess of GV galaxies in the outskirts of systems—consistent across simulations, mock data, and SDSS observations—highlights the importance of the environment in regulating star formation during the transition from the blue cloud to the red sequence.

\begin{acknowledgements}
This work has been partially supported with grants from Agencia Nacional de Promoción Científica y Tecnológica, the Consejo Nacional de Investigaciones Científicas y Técnicas (CONICET, PIP-2022-11220210100064CO), from the Agencia Nacional de Promoción de la Investigación, el Desarrollo Tecnológico y la Innovación de la República Argentina (PICT-2020-3690), from the Secretaría de Ciencia y Tecnología de la Universidad Nacional de Córdoba (SECYT-UNC, Res. 258/53). FR would like to acknowledge support from the ICTP through the Junior Associates Programme 2023-2028. 

The TNG 300-1 simulations used in this work are part of the IllustrisTNG project, which were run on the HazelHen Cray XC40 system at the High-Performance Computing Center Stuttgart, as part of the GCS-ILLU project of the Gauss Centers for Supercomputing (GCS).
\end{acknowledgements}

\bibliographystyle{aa} 
\bibliography{biblio} 

\begin{appendix}

\section{Complementary figures}

\begin{figure}[!h]
\centering
{\includegraphics[width=0.5\textwidth]{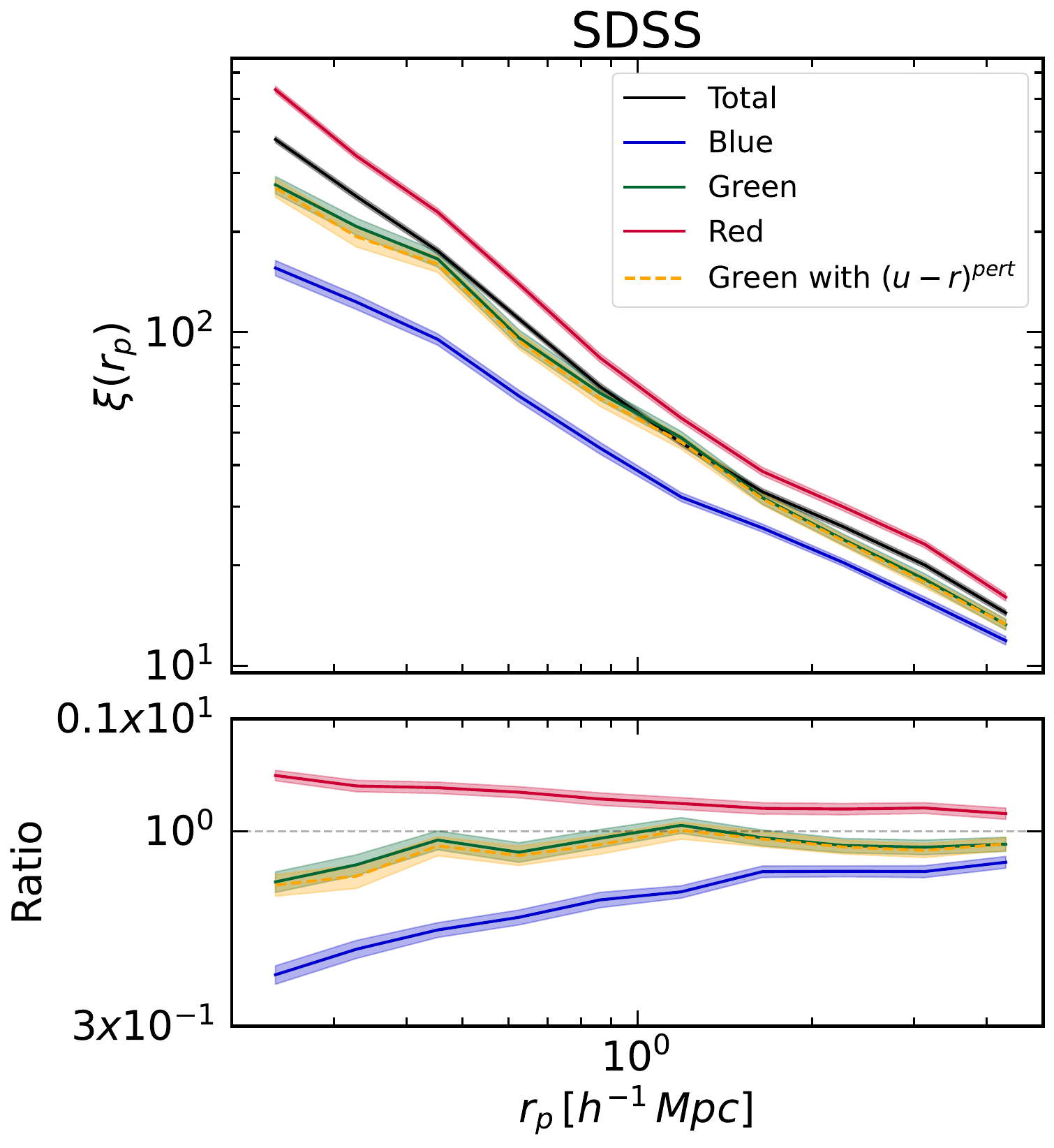}
\caption{\label{fig:ccf-sdss-color-pert} Upper panel: CCFs measured from the SDSS DR18 sample. The black curve shows the full sample, with all galaxies used as tracers and all galaxy groups used as centres. The blue, green, and red curves correspond to the CCFs of the respective colour subsamples. The orange dashed curve shows the CCF for the galaxy subsample of green galaxies constructed by perturbed $(u-r)$ colour applied to the CMD as mentioned in the Sect. \ref{sec:sample_obs}. Lower panels: Ratios of the CCFs of the red, green, blue, and orange subsamples to that of the total sample. Shaded regions indicate uncertainties estimated via jackknife resampling.}}
\end{figure} 

\begin{figure}[!h]
\centering
{\includegraphics[width=0.5\textwidth]{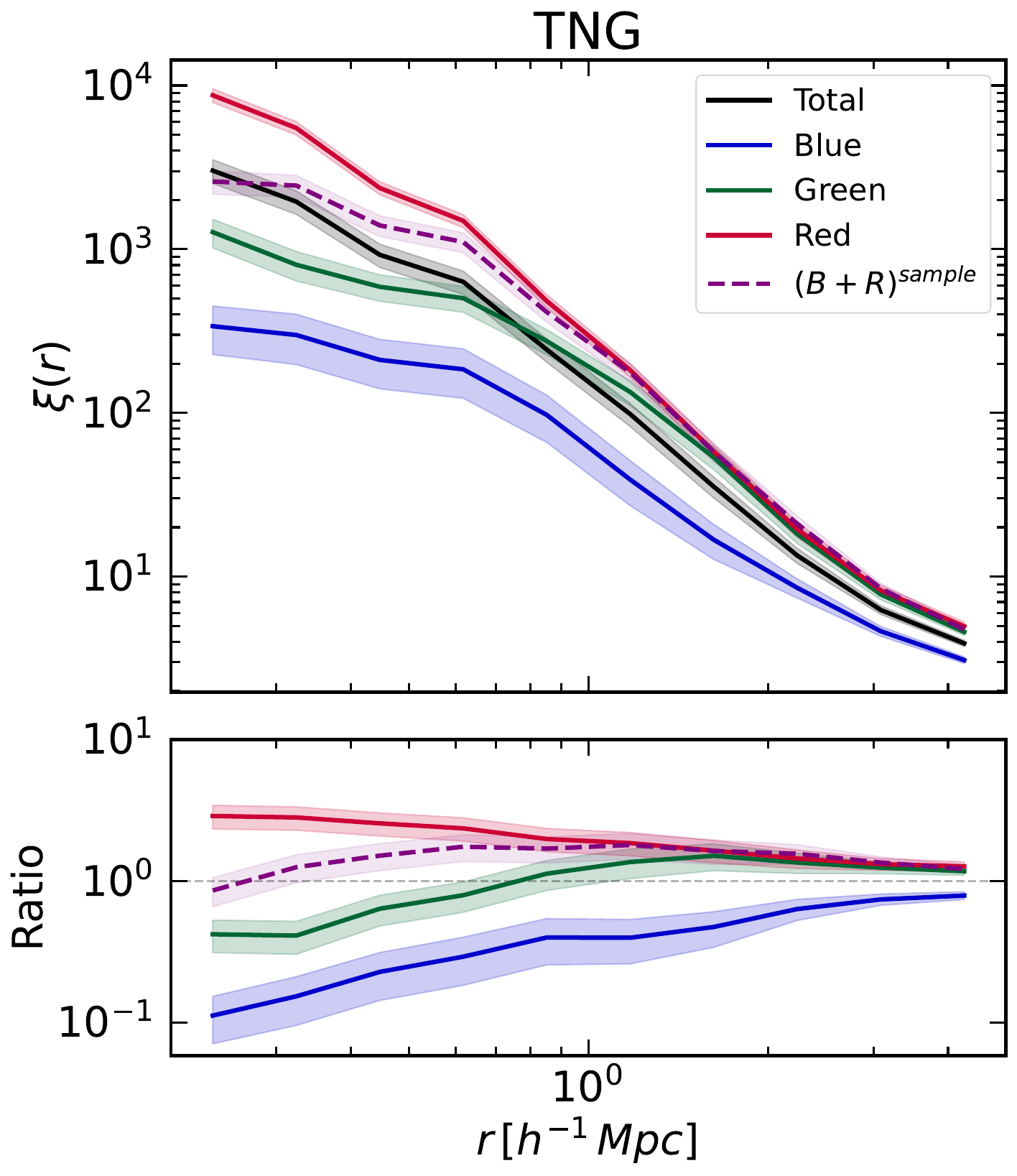}
\caption{\label{fig:ccf-tng-ap} Upper panel: CCFs measured from the IllustrisTNG300-1 simulation. The black curve shows the full sample. The blue, green, and red curves correspond to the CCFs of the respective colour subsamples. The violet dashed curve shows the CCF for the (B+R) sample selected as explained in Sect. \ref{sec:results}. Lower panels: Ratios of the CCFs of the red, green, blue, and violet subsamples to that of the total sample. Shaded regions indicate uncertainties estimated via jackknife resampling.}}
\end{figure} 

\begin{figure}[!h]
\centering
{\includegraphics[width=0.5\textwidth]{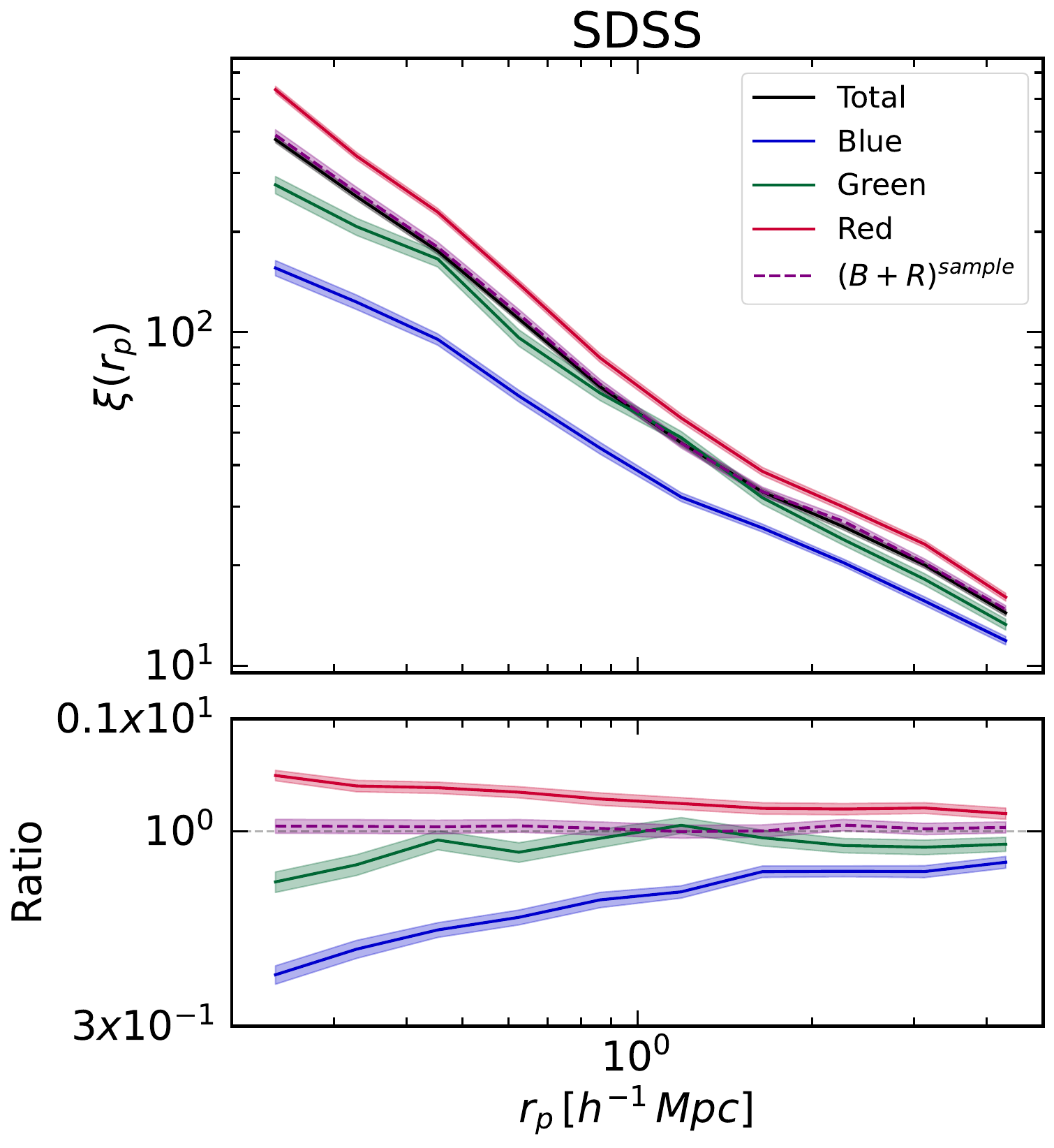}
\caption{\label{fig:ccf-sdss-ap} Same as Fig. \ref{fig:ccf-tng-ap} for the SDSS DR18 sample.}}
\end{figure}

\end{appendix}

\end{document}